\documentclass {mn2e}

\usepackage{graphicx}
\usepackage{epsfig} 

\def\gap{\;\rlap{\lower 2.5pt
 \hbox{$\sim$}}\raise 1.5pt\hbox{$>$}\;}
\def\lap{\;\rlap{\lower 2.5pt
   \hbox{$\sim$}}\raise 1.5pt\hbox{$<$}\;}
\def\gsim{\;\rlap{\lower 2.5pt
 \hbox{$\sim$}}\raise 1.5pt\hbox{$>$}\;}
\def\lsim{\;\rlap{\lower 2.5pt
 \hbox{$\sim$}}\raise 1.5pt\hbox{$<$}\;}
\def\msun{{\rm\,M_\odot}}

\def\cm{{\rm\,cm}}
\def\sec{{\rm\,s}}

\def\sr{{\rm\,sr}}

\def\cm{{\rm\,cm}}

\def\kpc{{\rm\,kpc}}

\def\km{{\rm\,km}}
\def\m{{\rm\,m}}
\def\GeV{{\rm\,GeV}}
\def\TeV{{\rm\,TeV}}
\def\sec{{\rm\,s}}
\def\sr{{\rm\,sr}}

\def\spose#1{\hbox to 0pt{#1\hss}}
\def\lta{\mathrel{\spose{\lower 3pt\hbox{$\mathchar''218$}}
     \raise 2.0pt\hbox{$\mathchar''13C$}}}
\def\gta{\mathrel{\spose{\lower 3pt\hbox{$\mathchar''218$}}
     \raise 2.0pt\hbox{$\mathchar''13E$}}}
\newcommand{\beq}{\begin{equation}}
\newcommand{\eeq}{\end{equation}}
\newcommand{\be}{\begin{equation}}
\newcommand{\ee}{\end{equation}}

\newcommand{\ls}{\mathrel{\raise1.16pt\hbox{$<$}\kern-7.0pt 
\lower3.06pt\hbox{{$\scriptstyle \sim$}}}}         
\newcommand{\gs}{\mathrel{\raise1.16pt\hbox{$>$}\kern-7.0pt 
\lower3.06pt\hbox{{$\scriptstyle \sim$}}}}         

\long\def\comment#1{}

\def\phcms{$\rm ph\;cm^{-2}\;s^{-1}$}

\def\msun{M_{\odot}}
\def\fun#1#2{\lower3.6pt\vbox{\baselineskip0pt\lineskip.9pt
  \ialign{$\mathsurround=0pt#1\hfil##\hfil$\crcr#2\crcr\sim\crcr}}}
\def\lap{\mathrel{\mathpalette\fun <}}
\def\gap{\mathrel{\mathpalette\fun >}}
\newcommand{\ba}{\begin{eqnarray}}
\newcommand{\ea}{\end{eqnarray}}

\title{Constraining the Sommerfeld enhancement with  Cherenkov telescope observations of  dwarf galaxies}

\author[L. Pieri, M. Lattanzi, J. Silk]
{Lidia Pieri$^{1}$, Massimiliano Lattanzi$^{2,3}$ 
and
Joseph Silk$^{2}$ \\
$^1$ Instiitute d'Astrophysique de Paris, France \\
$^2$ Physics Department, University of Oxford, UK  \\
$^3$ Istituto Nazionale di Fisica Nucleare, Rome, Italy
}

\begin{document}
\maketitle
\begin{abstract}
The presence of dark matter in the halo of our galaxy could be revealed through indirect
detection of  annihilation products. 
Dark matter annihilation is one of the possible interpretations of the recent measured excesses in positron and electron fluxes, once boost factors of the order of $10^3$ or more are taken into account. 
Such  boost factors are actually achievable through the velocity-dependent Sommerfeld enhancement of the annihilation cross-section.
Here we study the expected $\gamma$-ray flux from two local dwarf galaxies for which 
air Cerenkov measurements are available, namely Draco and Sagittarius.
We use velocity dispersion measurements to model the dark matter halos of the dwarfs, and the results of numerical simulations to model the presence of an associated population of subhalos.
We incorporate the Sommerfeld enhancement of the annihilation cross-section.
We compare our predictions with observations of Draco and Sagittarius performed by MAGIC and HESS, respectively. We also compare our results with the sensitivities of Fermi and of the future Cherenkov Telescope Array.
We find that the boost factor due to the Sommerfeld enhancement is already constrained by the MAGIC and HESS data, with enhancements greater than $5 \times 10^4$ being excluded. 
While Fermi will not be able to detect $\gamma$-rays from the dwarf galaxies s even with the most optimistic  Sommerfeld effect, we show that the Cherenkov Telescope Array will be able to test enhancements greater than $1.5 \times 10^3$.
\end{abstract}

\noindent

\section{Introduction}
Detection of a rise in the high energy cosmic ray $e^+$ fraction by the PAMELA satellite experiment \cite{PAMELA} and of a peak in the  $e^+  + e^-$ flux by the ATIC balloon experiment \cite{chan} has stimulated considerable recent 
theoretical activity in indirect detection signatures of particle dark matter via annihilations of the SUSY LSP and other massive particle candidates 
\cite{Hooper:2009fj,deBoer:2009rg,Grajek:2008pg,Hooper:2008kv,Liu:2008ci,Cholis:2008wq,Donato:2008jk,Cirelli:2008jk,Bergstrom:2008gr}. 
Several hurdles must be surmounted  if these signals are to be associated with dark matter annihilations,  Firstly, a high boost factor $(10^3-10^4)$ is needed within a kiloparsec of the solar circle \cite{Cirelli08}. Secondly, the boost factor must be suppressed in the inner galaxy to avoid excessive gamma ray emission, synchrotron radio emission, and $\bar p $ production \cite{Bertone08}.  Thirdly, the annihilation channels must be largely lepton--dominated \cite{Cirelli:2008pk}.   

The last of these requirements is addressed  in various particle physics models for the dark matter candidate \cite{Cirelli:2008pk}. Here we explore the implications of the first two of these requirements. The first is resolved via the Sommerfeld enhancement of the annihilation cross-section in local dark halo substructure  \cite{Lattanzi:2008qa,ArkaniHamed:2008qn}.This boost is especially relevant  on scales that are hitherto unresolved by numerical simulations \cite{Springel:2008by}. Indeed, the second requirement can be understood  because the unresolved substructures that dominates the local boost  are tidally disrupted in the inner  galaxy \cite{Lattanzi:2008qa}. However it is essential to test these assumptions, especially because they go beyond the range of current simulations.

If the Sommerfeld enhancement is indeed dominated by unresolved cold halo substructures, we show here that gamma rays from nearby dark matter--dominated dwarf galaxies may be detectable with experiments such as MAGIC and HESS. Moreover these experiments but in particular the future CTA, may allow imaging of both the smooth and subhalo components of nearby dwarfs such as Draco and Sagitarius.

\section{$\gamma$-ray flux from Dark Matter annihilation in Draco and Sagittarius}

The observed photon flux from DM annihilation can be written as

\begin{equation}
\frac{d \Phi_\gamma}{dE_\gamma}(M,E_\gamma, \psi, \theta) =
 \frac{1}{4 \pi} \frac{\sigma v }{2 m^2_\chi} \cdot 
\sum_{f} \frac{d N^f_\gamma}{d E_\gamma} B_f  \int_{V}  \frac{\rho_\chi^2(M,R)}{d^2} dV
\label{flussodef}
\end{equation}
where $M$ is the mass of the halo, $d$ the distance from the observer, $m_\chi$ denotes the dark matter particle mass and $d N^f_\gamma / dE_\gamma$ is the differential photon
spectrum per annihilation relative to the 
final state $f$, with branching ratio $B_f$. The volume integral refers to the line of sight and is defined by the angular resolution of the instrument $\theta$ and by the direction $\psi$ of observation. $R$ is the distance from the halo center. $\rho_\chi(M, c (M), R)$ is the dark matter density profile inside the halo, $c(M)$ being the concentration parameter of the halo, defined as the ratio between virial radius and scale radius and computed following the prescriptions of \cite{Bullock01} [B01].
The annihilation cross section $ \sigma v$ for a typical DM candidate is found to be appropriate for thermal relics that satisfy the cosmological constraints on the present abundance of dark matter in the universe. 

In cases where the Sommerfeld enhancement is present (see the next section for the details of the model) we replace the term $\sigma v $  in eq. \ref{flussodef} with
the velocity-dependent expression $\sigma v S(R,M)$
where the enhancement $S$ depends on the halo mass which in turn fixes the average velocity dispersion, and from the radial coordinate R inside the halo, which takes into account the features of the velocity dispersion curve that has lower values closer to the center of the galaxy hosted by the DM halo.
The highest values for $S$ are obtained for very high DM particle masses (of the order of a few TeV). In this mass region, the primordial annihilation cross-section can be as high as 
$10^{-26} \cm^3 \sec^ {-1}$, and we will use this upper value as a reference throughout our paper.

Including the Sommerfeld enhancement, Eq. \ref{flussodef}  will transform into
\begin{equation}
\Phi_\gamma(M,E_\gamma, \psi, \theta) \propto  \int \frac{S(M,R) \rho_\chi^2(M,c,R) }{d^2} dV
=  \Phi_S
\label{flussodefS}
\end{equation}

\subsection{The particle physics sector}
\begin{figure}
	\begin{center}
	\includegraphics[width=0.45\textwidth,  keepaspectratio]{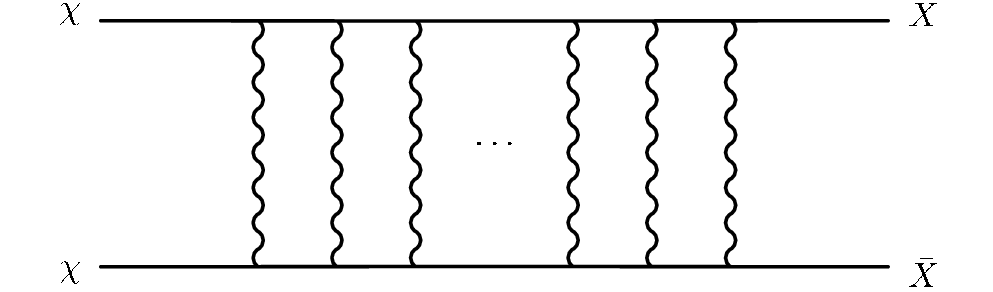}
\caption{Ladder diagram giving rise to the Sommerfeld enhancement for $\chi\chi\to X\bar X$ annihilation, via the exchange of gauge bosons. }
	\label{fig:somgen}
         \end{center}
\end{figure}

The dark matter annihilation cross section can be enhanced, with respect to its primordial value, in the presence of the so-called Sommerfeld effect.
This is a (non-relativistic) quantum effect occurring when the slow-moving annihilating particles interact through a potential \cite{somm31}. From the point of view of quantum field theory, the Sommerfeld effect is due to the resummation of ladder diagrams like the one shown in Fig. \ref{fig:somgen}.

The idea that the gamma-ray flux from dark matter annihilations can be enhanced in this way was first proposed in a pioneering paper by \cite{hisano} (see also \cite{Hisano05}). Recently, the possibility of explaining the large boost factor required by PAMELA using this mechanism has stimulated several studies of this effect (see for example \cite{Cirelli:2007xd,MarchRussell:2008yu,ArkaniHamed:2008qn,Pospelov08,Lattanzi:2008qa,MarchRussell:2008tu}). In the following, we will briefly summarize  some basic properties of the Sommerfeld enhancement.

In the presence of the enhancement, the effective s-wave annihilation cross section times velocity can be written as:
\beq
\sigma v = S(\beta, m_\chi) \left(\sigma v \right)_0,
\eeq
where $(\sigma v)_0$ denotes the tree level s-wave annihilation cross section, and we have explicitly indicated the dependence of the Sommerfeld enhancement $S$ on the particle mass $m_\chi$ and velocity $\beta=v/c$.
The Sommerfeld enhancement can be obtained solving the $\ell=0$ Schr\"odinger equation for the reduced two-body wave function $\psi(r)$:
\beq
\left(\frac{1}{m_\chi}\frac{d^2}{dr^2}-V(r)\right)\psi(r) = -m_\chi\beta^2\psi(r),
\eeq
with the boundary condition $\psi'(\infty)/\psi(\infty)=im_\chi\beta$. The Sommerfeld factor $S$ is then given by $S=|\psi(0)/\psi(\infty)|^2$.

In the following we will consider particles interacting through a Yukawa-like potential:
\beq
V(r)=-\frac{\alpha}{r} e^{-Mr},
\eeq
where $M$ is the mass of the exchange boson mediating the interaction. When we will have to specify numerical values for the interaction parameters,
we will take $\alpha=1/30$ and $M=90\,\GeV$, corresponding to particles interacting through the exchange of a $Z$ boson.

\begin{figure}
	\begin{center}	
	\epsfig{file=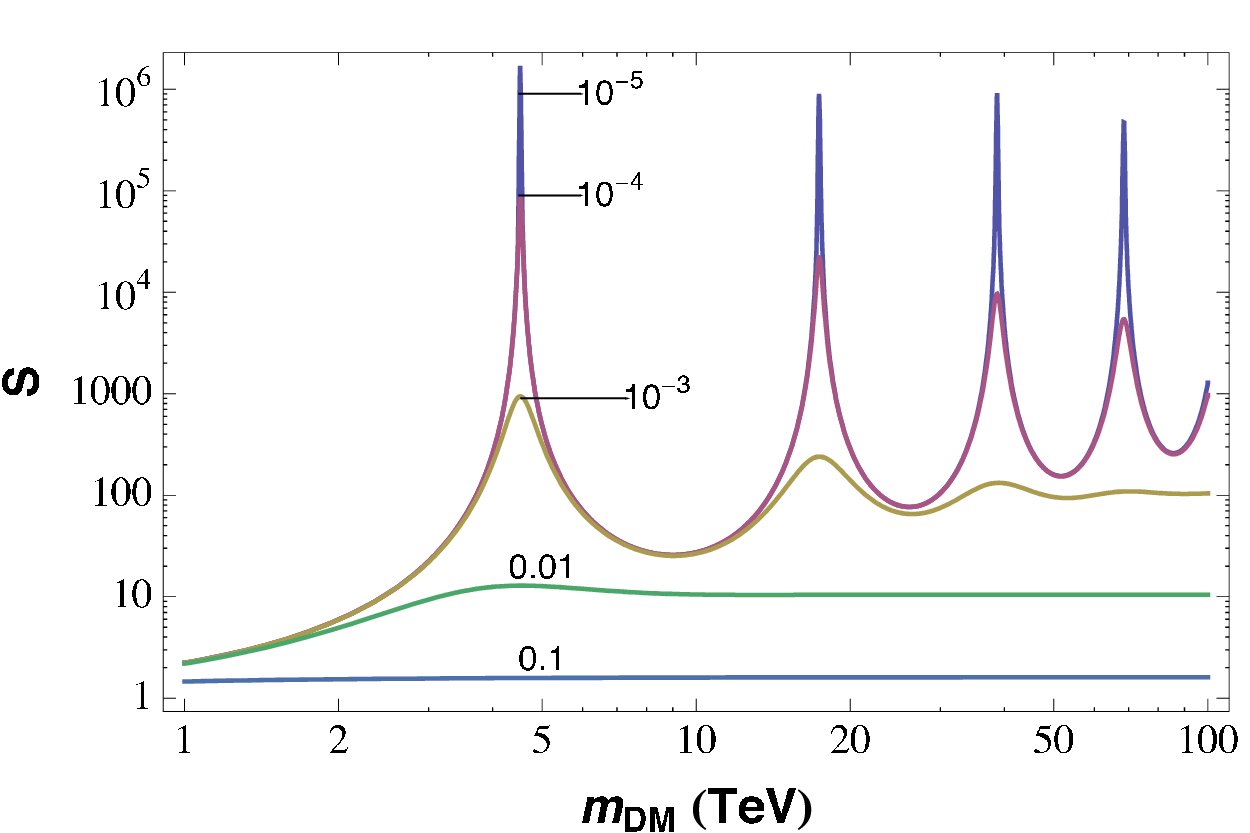,width=0.45\textwidth}
	\epsfig{file=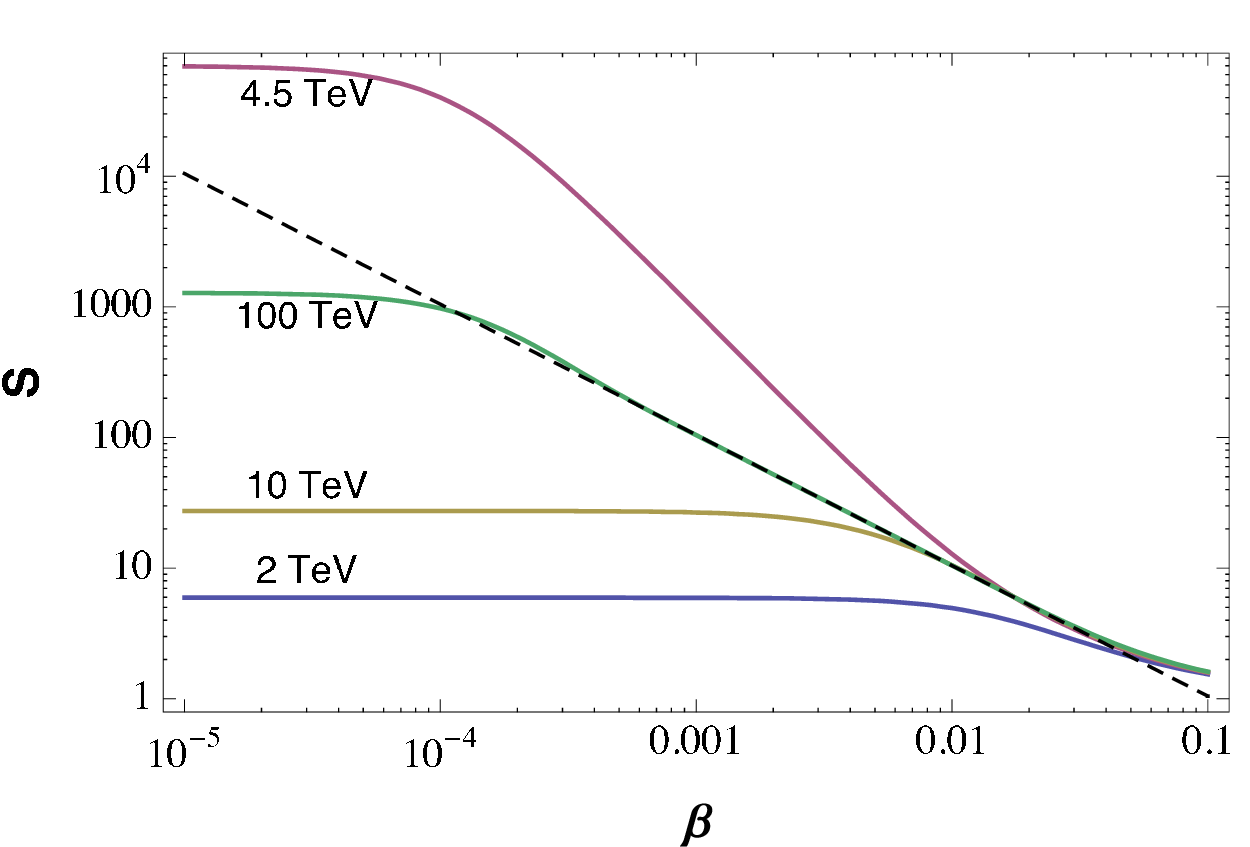,width=0.45\textwidth}
	\label{fig:som1}
	\caption{Top panel: Sommerfeld enhancement $S$ as a function of the dark matter particle mass $\m_\chi$, for different values of the particle velocity. Bottom panel: Sommerfeld enhancement $S$ as a function of the particle velocity $\beta$ for different values of the dark matter mass. The black dashed line shows the $1/v$ behaviour that is expected in the intermediate velocity range.}
         \end{center}
\end{figure}

The Sommerfeld enhancement is effective in the low-velocity regime, and disappears ($S=1$) in the limit $\beta\to 1$. In general, one can distinguish two distinct behaviours,  resonant and  non-resonant, depending on the value of the annihilating particle mass.
In the non-resonant case, the cross section is enhanced for $\beta < \alpha$: $S\simeq \pi \alpha/\beta$ up to a saturation value, roughly given by $S_{\mathrm{max}} \sim 6\alpha M/m_\chi$. This value occurs for $\beta \sim 0.5 M/m_\chi$.
In the resonant case, occurring for particular values of the mass of the annihilating particle, the 
cross-section follows the non-resonant behaviour until $\beta\simeq\sqrt{\alpha M/m}$; below this critical value, the enhancement grows like $1/\beta^2$ before saturating. The Sommerfeld boost can then reach very large values. These different behaviours can be observed in Fig.  \ref{fig:som1}.

We consider a WIMP dark matter candidate that annihilates mainly to weak gauge bosons. If the dark matter is a Majorana particle, such as for example the supersymmetric neutralino, its annihilation into fermionic final states is helicity-suppressed by a factor $(m_f/m_{\chi})^2$ . For a dark matter particle in the 1 to 10 TeV range, this is a factor $10^{-2}\div10^{-4}$ even for the heaviest possible final state, i.e. the top quark. However, for completeness we have also considered the heavy quark and lepton annihilation channels. The differential photon spectra per annihilation $dN^f_\gamma/dE_\gamma$ for the various final states have been computed using PYTHIA \cite{Sjostrand:2000wi}, including also the contribution from final state radiation.

\begin{figure}
	\begin{center}
	\epsfig{file=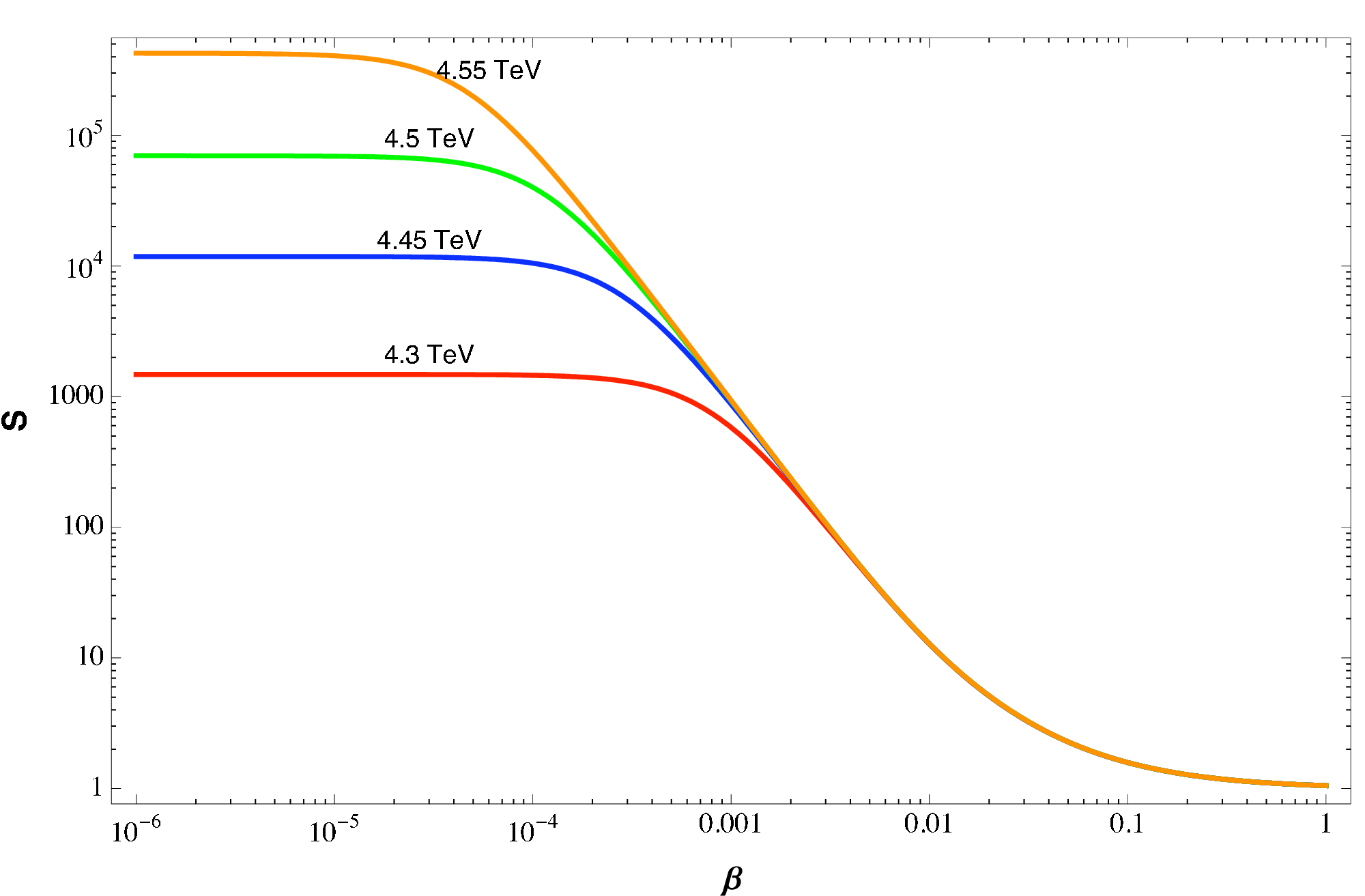,width=0.45\textwidth}
	\label{fig:som-res}
	\caption{Sommerfeld enhancement $S$ as a function of the particle velocity $\beta$ for different values of the dark matter mass close to the resonance.}
         \end{center}
\end{figure}

We consider values of the particle mass very close to the first resonance that can be seen in the top panel of Fig. \ref{fig:som1}, i.e $m_\chi\simeq 4.5 \TeV$, in order to obtain the large boost factor required to explain the positron excess. In particular, we consider the following values for the mass of the particle: $m_\chi = (4.3,\,4.45,\,4.5,\,4.55\,\TeV)$. Being so close to the resonance, even a relatively small change in the mass of the particle can produce order of magnitude changes in the Sommerfeld boost. In fact, the maximum achievable boost goes from $S\simeq1.5\times10^3$ for $m_\chi=4.3\,\TeV$ to $S\simeq 4\times 10^5$ for $m_\chi=4.55\,\TeV$. We show the boost as a function of velocity in Fig. \ref{fig:som-res}; its main properties, i.e. the maximum value $S_{max}$ and the saturation velocity $\bar\beta$, are summarised in Table \ref{tab:tab1} for the different masses  As we show in the next sections, these large boost factors can be tested through Cherenkov telescope observations of dwarf galaxies.

\begin{table}
\begin{center}
\begin{tabular}{ccc}
Mass (TeV) & $S_{max}$ & $\bar\beta$ \\ 
\hline
4.3   & $1.5\times 10^3$ & $8.0\times 10^{-4}$ \\
4.45 & $1.2\times 10^4$ & $2.8\times 10^{-4}$ \\
4.5   & $7.0\times 10^4$ & $1.1\times 10^{-4}$ \\
4.55 & $4.2\times 10^5$ & $4.7\times 10^{-5}$ \\
\hline
\end{tabular}
\caption{Values of the maximum possible boost $S_{max}$ and of the saturation velocity $\bar\beta$, for different values of the dark matter particle mass.}
\label{tab:tab1}
\end{center}
\end{table}

\subsection{The astrophysical sector: smooth dark matter halo}
We deal with two dark matter-dominated dwarf galaxies of our Local Group, namely Draco and Sagittarius. We have chosen such galaxies because they have been observed by the Air Cherenkov Telescopes. Draco, which is visible from the northern hemisphere, has been observed by MAGIC \cite{MAGIC}, while the flux of $\gamma$-rays from Sagittarius, which is closer to the Galactic Centre and visible from the southern hemisphere, has been measured by HESS \cite{HESS}.
Both MAGIC and HESS did not observe any signal and therefore put upper limits on the $\gamma$-ray coming from these sources. \\

The Draco galaxy lies about 80 kpc away from us, at the zenith with respect to the Galactic Center. 
In order to model its dark matter halo, we use the density profile obtained by \cite{Walker2007}, who fitted the velocity dispersion measurements of its stellar population adopting a one-component King profile for the luminous component, and a Navarro, Frenk and White (NFW) profile \cite{Navarro1996,Navarro1997} with constant anisotropy parameter for the DM one.
We find that the data of \cite{Walker2007} are well fitted by a total initial virial mass of $5 \times 10^9 \msun$, and NFW scale parameters $r_s = 2.04 \kpc$ and $\rho_s = 0.82 \GeV \cm^{-3}$. The virial radius which encloses this initial mass is $\sim 45 \kpc$.
As confirmed by the recent high resolution N-body simulations {\it Aquarius}  \cite{Springel:2008by,Springel:2008cc} and {\it Via Lactea II} \cite{Diemand:2008in}, the satellites, or subhalos,
of our Galaxy suffer from external tidal stripping due to the interaction with the Milky Way.
To account for gravitational tides, we follow \cite{Hayashi} and assume that all the mass beyond the subhalo tidal radius is lost in a single orbit without affecting its central density profile. The tidal radius is defined as the distance from the subhalo center at which the tidal forces of the host potential equal the self-gravity of the subhalo. In the Roche limit, it is expressed as:
\begin{equation}
r_{tid}(r)= \left (\frac{M_{sub}}{2 M_{h}(<r)} \right)^{1/3} r 
\end{equation}
where r is the distance from the halo center, $M_{sub}$ the subhalo mass and $M_{h}(<r)$ the host halo mass enclosed in a sphere of radius r. \\
In our case, the host halo is the Milky Way, which we describe with an NFW profile ($M=1.8 \times
10^ {12} \msun$, $r_s = 23.3 \kpc$, $\rho_s = 0.28 GeV \cm^ {-3} $ ). \\
At the distance of Draco, we find $r_ {tid} = 11.5 \kpc$.
We note that the condition $r_ {tid} > r_s$ holds, which guarantees that the binding energy is negative and the system is not dispersed by tides. \\

The Sagittarius dwarf galaxy is located at a distance of about 24 kpc from us, at low latitudes. Its vicinity to the Galactic Center causes significant tidal stripping due to the interaction with the gravitational potential of the Milky Way. Yet the surviving stellar component suggests that its inner dark matter halo also survives.
The observations show that Sagittarius  is indeed dark matter-dominated with a central stellar velocity dispersion of about 10 km s$^{-1}$ \cite{Ibata97}, similar to the one observed in Draco. 
We modeled the inner regions of the DM halo of the Sagittarius dwarf with the same scale parameters as Draco (see also \cite{Evans2004}) obtained assuming a NFW mass density profile and an initial mass of $M = 5 \times 10^9$ M$_\odot$.  \\
At the distance of Sagittarius, the tidal radius is $r_ {tid} = 5.3 \kpc$, still larger than the scale radius. \\

The astrophysical contribution to the $\gamma$-ray emission from the smooth component of a DM halo can be written as the volume integral
\begin{equation}
\Phi_S(M, d, \psi, \Delta \Omega) \propto \int \int \int_{V}  
d \phi d \theta d\lambda  \left [ \frac{S(M,R) \rho_{\chi}^2(M,R)} {\lambda^{2}}\right] \, 
\label{singlehalophicosmo}
\end{equation}
where  $\lambda$ is the line-of-sight coordinate,  $\Delta \Omega$ the solid angle corresponding to the angular resolution $\theta$ of the instrument, and $\psi$ the angle of view from the GC; in the case of ACTs,  $\theta = 0.1^\circ$ and  $\Delta \Omega \sim 10^{-5} \sr$. 
The integral along the line-of-sight will be different from zero only in the interval 
[$d - r_{tid}, d + r_{tid}]$. \\

In the case of the dwarf galaxies, their mass and therefore the masses of the sub-subhalos lie in the region at low $\beta$ where the Sommerfeld enhancement saturates. This is true for every DM mass except for the one which lies closest to the resonance (in our model, $m_{DM} = 4.55 \TeV$). In this case, however, the radial dependence of the enhancement produces a variation of a few percent. As a first approximation, we can write 
\begin{equation}
\Phi_S = S \times \Phi^{\rm cosmo} \
\end{equation} 
where the latter term is just the astrophysical contribution to the $\gamma$-ray flux, without taking into account any enhancement coming from particle physics. We will however compute our prediction according to Eq. \ref{singlehalophicosmo} in the case of $m_{DM} = 4.55 \TeV$, when comparing our results to the existing data.

The result of the computation of $\Phi^{\rm cosmo}$ for Draco and Sagittarius are depicted in Fig.\ref{draco} as a function of $\psi$.

\subsection{The astrophysical sector: substructures}
Both the  {\it Aquarius}  and the  {\it Via Lactea II} simulations have succeeded in observing 
sub-subhalos, that is to say substructures inside the satellites of the Milky Way-like host halo. 

We  will therefore populate the dwarf galaxies with sub-subhaloes with masses as small as $10^{-6} M_{\odot}$, which correspond  to the typical Jeans mass for  a generic CDM weakly
interacting massive  particle (WIMP) particle  with $m_{\rm DM}  = 100 \GeV$  \cite{ghs04,ghs05}.  Such  a  minimum mass may vary between $10^{-12}$  and $10^{-4} M_{\odot}$ depending on the particle physics \cite{profumo}.

We follow the results of  {\it Via Lactea II} to model the population of sub-substructures.
We then adopt a sub-substructure mass function at z=0
\begin{equation}  
{\rm d} n(M_{sub})/{\rm d ln}(M_{sub}) \propto M_{sub}^{-1},
\label{mass}  
\end{equation}  
and we model the radial distribution following \cite{kuhlen}:
\begin{equation}  
{\rm d} n(R)/{\rm d}(R) \propto   \frac{1}{(1 + R/r_s^{h})^{2}},  
\label{radius}  
\end{equation}  
where $r_s^{h}$ is the scale radius of the host halo and $R$ is the radial coordinate inside the host halo.

We normalize the subhalo distribution function $\rho_{sh}(M_h,M_{sub},R)$ such that 10 \% of the mass of the host halo ($M_h$) before the tidal stripping is distributed in substructures with masses between $10^{-5} M_{h}$  and $10^{-2} M_{h}$. 

The resulting distribution function is
\begin{equation}  
\rho_{sh}(M_{h},M_{sub},R) =  \frac{4.13 \times 10^{4}}{M_{sub}^{2}}  \frac{1}{(1 + R/r_s^{h})^{2}}
\msun \kpc^{-3}
\label{rhosh}  
\end{equation} 
As a second step, we cut off all the subhalos which lie beyond $r_{tid}$.
This is indeed an upper value for the number of surviving sub-subhaloes, since we are  not considering here the fifty percent of the subhalos that exit the virial radius of the  parent halo during their
first  orbit \cite{bepi} and are therefore  dispersed into the halo of the Milky Way. \\

In the case of Draco, we end up with about 16\% of 
the present Draco mass (inside $r_{tid}$) condensed in $\sim 1.2 \times 10^{13}$ sub-subhalos with 
masses in the range [$10^{-6},10^{7}] \msun$. \\

As far as Sagittarius is concerned, we get $\sim 3.3 \times 10^{12}$ sub-subhalos, accounting for 
4 percent of the total bound mass. As expected, the tidal disruption in this case is more efficient and 
sweeps away most of the substructures. \\

The contribution of such a population of sub-substructures to the annihilation signal can be 
written as \cite{PBB8}:
$$ 
\Phi^{\rm cosmo}(M_{h}, d, \psi, \Delta \Omega) \propto \int_{M_ {sub}} d M_ {sub} \int_c d c \int \int_{\Delta \Omega} 
d \theta d \phi 
$$
\begin{equation}
\int_{\lambda}  d\lambda [ \rho_{sh}(M_{h},M_{sub},R) P(c)  \Phi^{\rm cosmo}_{sh}(M,c(M,R),d,\psi, \Delta \Omega) ]
\label{smoothphicosmo}
\end{equation}
where the contribution from each sub-subhalo ($\Phi^{\rm cosmo}_{sh}$) is convolved with its distribution function ($\rho_{sh}$).
$P(c)$ is the lognormal distribution of the concentration parameter with dispersion 
$\sigma_c$ = 0.24 \cite{Bullock01} and mean value $ \bar{c}$:
\begin{equation}
P(\bar{c},c) = \frac{1}{\sqrt{2 \pi} \sigma_c c} \, 
e^{- \left ( \frac{\ln(c)-\ln(\bar{c})} {\sqrt{2} \sigma_c} \right )^2}.
\end{equation}
Again, the integral along the line-of-sight will be different from zero only in the interval 
[$d - r_{tid}, d + r_{tid}]$.

For each sub-substructure, we use an NFW density profile whose concentration
parameter $c(M,R)$ depends on its mass and on its position inside the host halo, according to the results of  {\it Via Lactea II} :
\begin{equation}
c(M,R)=c_{B01}(M)\left (\frac{R}{R_{\rm vir,h}} \right )^{-0.286}.
\end{equation}
The mass dependence $c_{B01}(M)$ is taken from B01 and extrapolated with a double power law down to the smallest masses, and $R_{\rm vir,h}$ is the virial radius of the host halo.

We numerically integrate Eq. \ref{smoothphicosmo} and estimate the 
contribution to  $\Phi^{\rm cosmo}$ from the sub-substructures in a $10^{-5} \sr$ 
solid angle along the direction $\psi$. \\

\begin{figure}
\epsfig{file=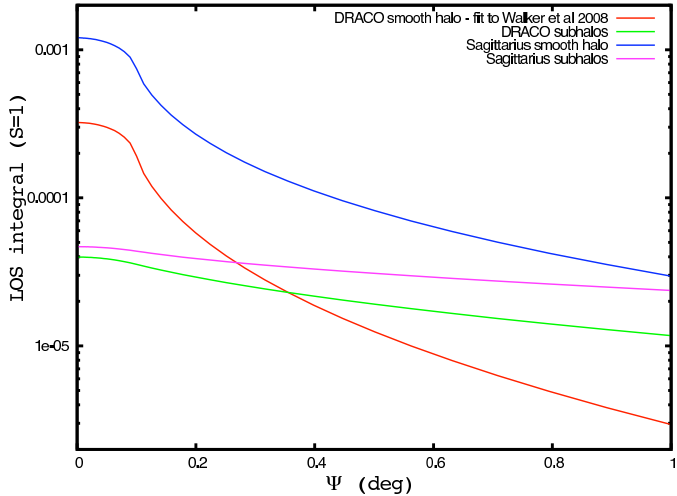,width=0.5\textwidth}
\caption{
$\Phi_{cosmo}$ as a function of the angle of view $\psi$ from the centre of halo, computed in the case of Draco and Sagittarius, for the smooth halo and from the subhalo population.}
\label{draco}
\end{figure}

The result of the computation of $\Phi^{\rm cosmo}$ for the subhalo population of Draco and Sagittarius are depicted in Fig.\ref{draco} as a function of $\psi$.
The presence of sub-subhalos is unimportant in the case of Sagittarius.
In the case of Draco, it becomes relevant only away from its center, where gives anyway a flux which is one order of magnitude smaller.

\subsection{Comparison with the experimental data}
The MAGIC and HESS ACTs have put upper limits on the $\gamma$-ray fluxes from Draco and Sagittarius, respectively. 
The upper limit for Draco integrated over energies above 140 GeV is $10^{-11}$ \phcms.
In the case of Sagittarius, this limit is $3.6 \times 10^{-12}$ \phcms, integrated above 250 GeV. \\

\begin{figure}
\epsfig{file=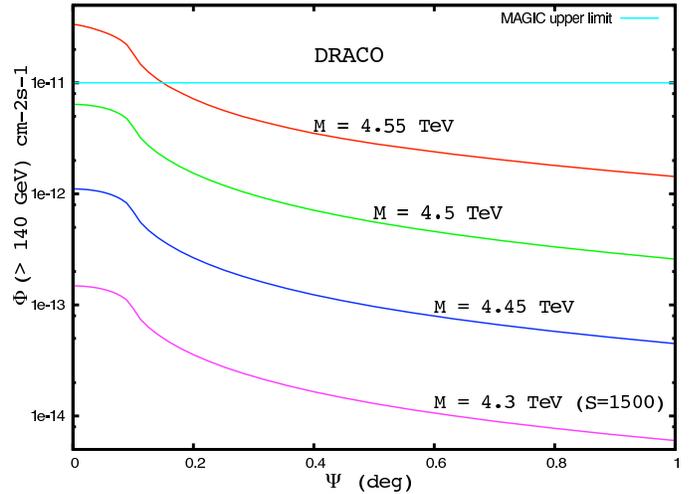,width=0.5\textwidth}
\caption{
Expected $\gamma$-ray flux above 140 GeV as a function of the angle of view $\psi$ from the centre of Draco.}
\label{magic}
\end{figure}

\begin{figure}
\epsfig{file=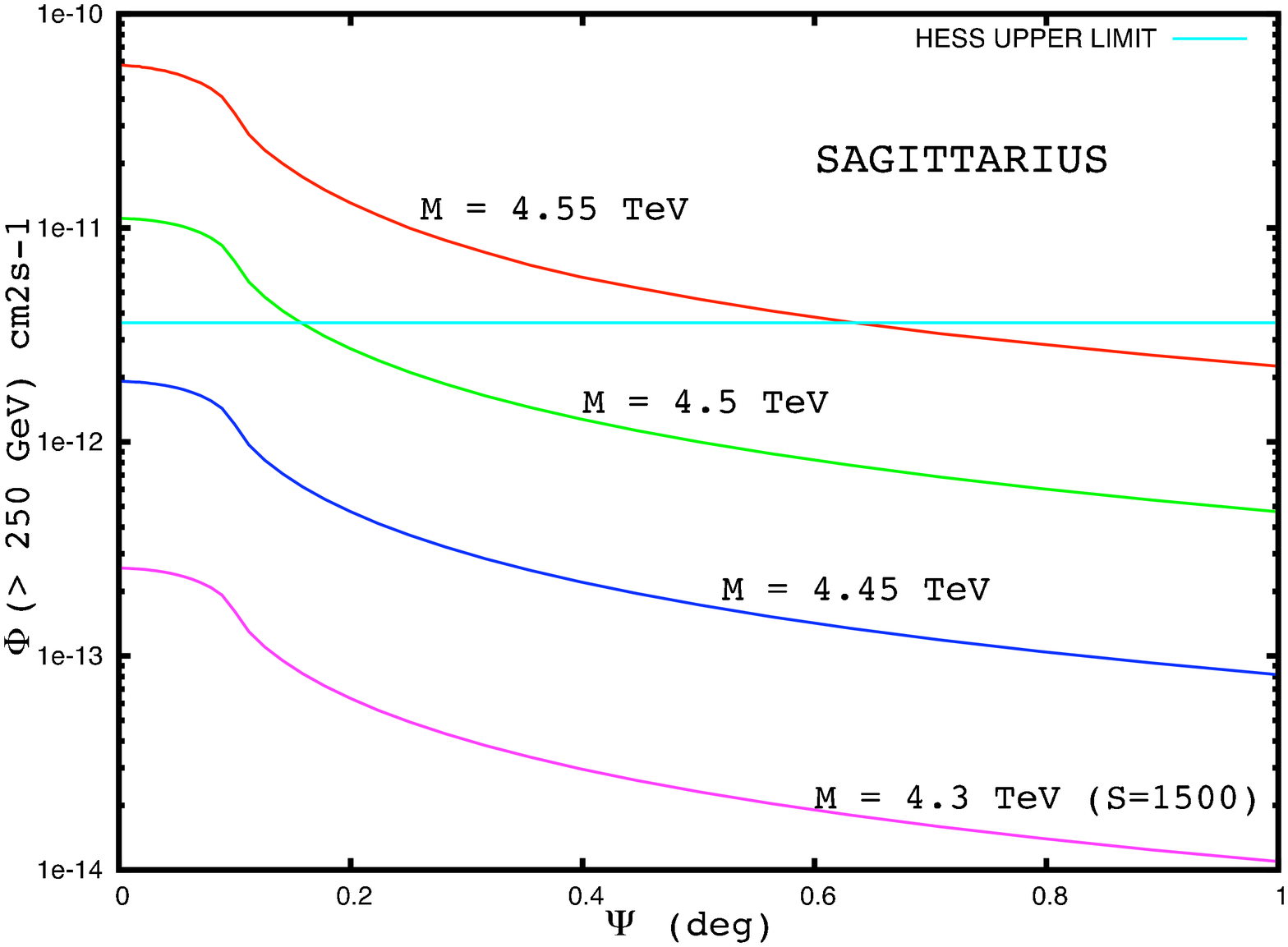,width=0.5\textwidth}
\caption{
Expected $\gamma$-ray flux above 250 GeV as a function of the angle of view $\psi$ from the centre of Sagittarius.
}
\label{hess}
\end{figure}

In Fig. \ref{magic} and \ref{hess} we compare these values with the prediction of the $\gamma$-ray flux from DM annihilations. We compute the flux for the DM mass values  give the greatest Sommerfeld enhancement, following eqs. \ref{flussodef} and \ref{flussodefS}. 
We can observe that the data from both MAGIC and HESS already exclude the greatest values for the enhancement. The most stringent limit is given by HESS, which constrains $S$ to be smaller than $5 \times 10^4$. \\

We have repeated our computation in the case of photon energies greater than 1 GeV, to compare with the sensitivity of Fermi to point sources \cite{Baltz2008}. We find that the detection of dwarf galaxies with Fermi is out of the range of experimental feasibility (see also \cite{Petal08}), even in the serendipitous case where  DM particles could have the enhancement necessary to produce the excess in electrons and positrons. This means that the ACTs 
provide the only possibility for discovering a possible source of $\gamma$-rays from DM annihilations in the
dwarfs, in the scenario where DM may be responsible for the positron excess. \\

\begin{figure}
\epsfig{file=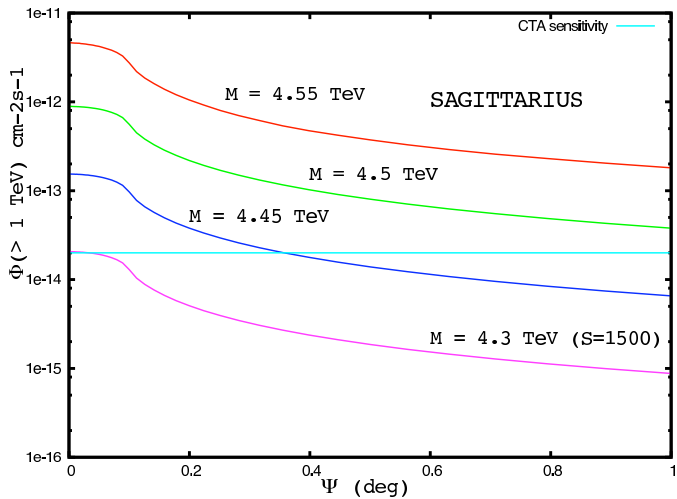,width=0.5\textwidth}
\caption{
Expected $\gamma$-ray flux above 1 TeV as a function of the angle of view $\psi$ from the centre of Sagittarius.}
\label{cta}
\end{figure}

We have therefore compared our predictions above 1 TeV with the expected sensitivity of the next-generation Cherenkov Telescope Array (CTA). The result is shown in Fig. \ref{cta}. If CTA will be built in the southern hemisphere, it will be able to test the enhancement down to the value of $1.5 \times 10^3$. In the case of no discovery, this means that the mechanism producing the electron-positron excess does not come from annihilating DM, since the allowed boost factors would be too low to explain the excess.

The results discussed so far have been obtained considering a dark matter particle of mass $m_\chi \simeq 4.5$ TeV annihilating exclusively into gauge bosons. Consideration  instead of heavy quarks or leptons as possible final states changes the predicted fluxes by factors of order unity, thus leaving our conclusions basically unchanged. In particular, a particle that annihilates only to heavy quarks would produce a flux 1.6-1.7 times larger than that shown in the figures, for all experiments. The limits on the Sommerfeld boost would then be proportionally tighter. In the case of a particle annihilating to $\tau$ leptons, the change in the flux depends on the energy threshold: for MAGIC, HESS and CTA it is respectively 0.5, 0.8, and 3.8 times the flux from the gauge boson channel.

\subsubsection{Constraints from the HESS observation of the Galactic Center source}
The HESS telescope has extensively observed the Galactic Center (GC) source, measuring an integrated flux above 1 TeV of $\Phi (> 1 \TeV) = 1.87 \times 10^{-12} \ ph \cm^{-2} \sec^{-1}$ in  2003 and 2004 \cite{hessgc}. \\

The  {\it Via Lactea II} and  {\it Aquarius} fits to the inner region of the Milky Way are not conclusive. 
Apart from the fact that the simulations do not include baryons which may play an important role at the GC, they disagree on the central slope which better fits the data.
While  {\it Aquarius} is better fitted with an Einasto profile with $\alpha = 0.21$, $r_s = 20 \kpc$ and 
$\rho_s = 2.1 \times 10^6 \msun \kpc^ {-3}$ (model GC-A), {\it Via Lactea II} finds an inner slope of -1.24, i.e. steeper than the NFW one, with $r_s = 28.1 \kpc$ and $\rho_s = 3.5 \times 10^6 \msun \kpc^ {-3}$ (GC-B), although they allow a fit with an NFW profile assuming $r_s = 21 \kpc$ and $\rho_s = 8.1 \times 10^6 \msun \kpc^ {-3}$ (GC-C).
The line-of-sight integral for an angular resolution of $0.1^\circ$ varies from 0.022 (GC-A) to 0.167 (GC-B) to 3.11 (GC-C). The HESS measurement allows us to set upper limits on the possible contribution 
due to the particle physics sectors. That is to say, we may allow a maximum particle physics contribution ranging from $8.5 \times 10^{-11}$ (GC-A) to $6 \times 10^{-13}$ (GC-C). Without taking into account the Sommerfeld enhancement, the particle physics contribution to the flux is $\sim 10^{-14}$ (averaged over
the mass range that we have explored here, 4.3 TeV $< m_{DM} <$ 4.55 TeV). \\
Our conclusion is that the maximum enhancement due to the Sommerfeld effect ranges from 60 (GC-C)
through 1120 (GC-B) to 8500 (GC-A). \\ 
The Sommerfeld enhancement computed for $v_{rot} \sim 200 \km \sec^{-1}$ ranges from 2140 ($m_{DM}$ = 4.55 TeV) to 880 ($m_{DM}$ = 4.3 TeV) so that, in the  {\it Aquarius} model the upper limit does not exclude any DM mass, while in the NFW case there is still room left for a 4.3 TeV DM particle. \\ 
The previous estimates are competitive with HESS limits on Sagittarius, although suffering from the large uncertainty  about the Galactic Center physics. It is anyway remarkable that the observation of Sagittarius with the CTA could give better limits than the GC region, in models with no DM spike at the GC.

\section{Conclusions}
The excess in cosmic-ray positrons and electrons has motivated a wealth of theoretical efforts in order to be explained in terms of DM. In particular, the annihilation mechanism has been revised in the light of the Sommerfeld enhancement, a velocity-dependent effect. 
Such an effect is maximal in the dwarf galaxies and in their substructures. The enhancement actually saturates for DM halo masses smaller than the dwarf scale.
In this work, we have computed the expected $\gamma$-ray flux from the Draco and the Sagittarius dwarfs galaxies, for which observational data are available from the ACTs.
We have adapted the smooth halo density profile in order to fit the measurements of velocity dispersions, and we have modeled the presence of a sub-subhalo population inside the dwarfs according to the results of the most recent N-body simulations of a Milky-Way sized halo. 
We found that the measurements of MAGIC and HESS are able to constrain the enhancement 
and put an upper limit on it of $5 \times 10^4$.
We have shown that the future CTA would be able to test values of the Sommerfeld enhancement as small as $1.5 \times 10^3$. Since such small values could not explain the excess in positrons/electrons, this means that the CTA would be able to confirm or exclude the interpretation of the excess in terms of annihilating DM. Finally, we have shown that, in the case where annihilating DM is responsible for the excess, Fermi will not be able to observe any signal.

\end{document}